# Leveraging generative adversarial networks with spatially adaptive denormalization for multivariate stochastic seismic data inversion

Roberto Miele[1,2,*], Leonardo Azevedo[1]

[1]CERENA/DER, Instituto Superior Técnico, University of Lisbon, Av. Rovisco Pais 1, 1049-001, Lisbon, Portugal.
[2]Institute of Earth Sciences, University of Lausanne, Géopolis, 1015, Chavannes-près-Renens, Switzerland.
*Corresponding author e-mail:* roberto.miele@unil.ch

## Abstract

Probabilistic seismic inverse modeling often requires the prediction of both spatially correlated geological heterogeneities (e.g., facies) and continuous parameters (e.g., rock and elastic properties). Generative adversarial networks (GANs) provide an efficient training-image-based simulation framework capable of reproducing complex geological models with high accuracy and comparably low generative cost. However, their application in stochastic geophysical inversion for multivariate property prediction is limited, as representing multiple coupled properties requires large and unstable networks with high memory and training demands. A more recent variant of GANs with spatially adaptive denormalization (SPADE-GAN) enables the direct conditioning of facies spatial distributions on local probability maps. Leveraging on such features, an iterative geostatistical inversion algorithm is proposed, integrating a pre-trained SPADE-GAN with geostatistical simulation, for the prediction of facies and multiple correlated continuous properties from seismic data. The SPADE-GAN is trained to reproduce realistic facies geometries, while sequential stochastic co-simulation predicts the spatial variability of the facies-dependent continuous properties. At each iteration, a set of subsurface realizations is generated and used to compute synthetic seismic data. The realizations providing the highest similarity coefficient to the observed data are used to update the conditioning probability models of each subsurface property in the subsequent iteration. This approach is referred to as SPADE-GANInv. The method is demonstrated on both 2-D synthetic scenarios and field data, targeting the prediction of facies, porosity, and acoustic impedance from full-stack seismic data. Results show that the algorithm enables accurate multivariate prediction, mitigates the impact of biased prior data, and accommodates additional local conditioning such as well logs. The integration of SPADE-GAN with geostatistical modeling provides flexibility and adaptability across inversion grid sizes, spatial patterns, and subsurface parameters.

# 1. Introduction

Accurate environmental risk management and resource exploration rely on characterizing multiple (multivariate) subsurface geological, petrophysical, and elastic properties (e.g., Rubin and Hubbard, 2005; Grana et al., 2021). Spatial geological heterogeneities (e.g., facies) and correlated continuous rock and elastic properties (e.g., porosity and acoustic impedance) are often linked by complex and nonlinear relationships, represented by statistical or deterministic rock physics models (e.g., Avseth, 2010; Mavko et al., 2020). Because direct observations (e.g., well data) are sparse, subsurface characterization often relies on geophysical data, such as seismic reflection, which are related to the properties of interest through nonlinear physical relations (e.g., Grana et al., 2022b; Tylor-Jones and Azevedo, 2024). This involves solving an underdetermined, ill-posed inverse problem with non-unique solutions (Tarantola, 2005).

In multivariate scenarios, deterministic approaches are computationally advantageous, as they seek an optimal solution from an initial model, using efficient mathematical frameworks (e.g., Tarantola, 2005; Sen and Stoffa, 2013) but tend to yield smoothed, geologically unrealistic results with limited uncertainty quantification (e.g., Rubin and Hubbard, 2005; Linde et al., 2017; Tylor-Jones and Azevedo, 2024). Probabilistic inversion approaches, on the other hand, predict a distribution of possible solutions that account for the uncertainties in the inverse problem and integrate prior spatial models to constrain the solutions; the results are represented as a probability density function or as a set of realizations (Grana et al., 2022b). Bayesian inference (Tarantola, 2005) is a widely adopted approach, involving the prediction of the subsurface parameters' posterior distribution by means of sampling algorithms based on Monte Carlo Markov chains (MCMC; Mosegaard and Tarantola, 1995, 2014; Connolly and Hughes, 2016; Grana et al., 2019, 2022a; Amaya et al., 2021) or variational inference (Nawaz and Curtis, 2019; Nawaz et al., 2020; Levy et al., 2023; Miele et al., 2024). However, multivariate seismic inversion case studies are often characterized by very large computational demand, due to the required large prior models and extensive parameters exploration. In this context, linearized Bayesian inversion methods (e.g., Buland and Omre, 2003; Grana and Della Rossa, 2010; Grana et al., 2017; Guo et al., 2021) can efficiently derive analytical solutions for the posterior distribution but are limited in the integration of spatial information as prior constraint, and incur an underestimation of uncertainties.

Alternatively, stochastic optimization algorithms, such as simulated annealing and neighborhood algorithms (e.g., Sen and Stoffa, 1991; Sambridge, 1999; Coléou et al., 2005; Tang et al., 2024), probability perturbation methods (Caers and Hoffman, 2006; González et al., 2008; Azevedo et al., 2019, 2020a; Miele et al., 2022a, 2022b), and genetic algorithms (Sambridge and Drijkoningen, 1992; Curtis and Snieder, 1997; Jia et al., 2021), are generally designed to address nonlinear inverse problems with complex or discontinuous objectives. Among these, iterative geostatistical seismic inverse methodologies (Azevedo and Soares, 2017; Tylor-Jones and Azevedo, 2024) use geostatistical sequential simulation and co-simulation (Deutsch and Journel, 1992) to represent spatially correlated multivariate subsurface properties models, and perform their perturbation and update based on the observed seismic data. The geostatistical modeling of continuous properties honors spatial patterns based on two-point statistics (i.e., variogram models), while facies distributions are inferred through classification conditioned on the rock properties (e.g., Azevedo et al., 2015; Miele et al., 2022a). To account for prior spatial information in the facies domain, Azevedo et al. (2020a) proposed the use of one-dimensional Markov chains to model facies each seismic trace location, and condition facies-dependent geostatistical co-simulation of the continuous properties. However, both Markov chains and two-point-based geostatistical simulation methods are particularly limited in the reproduction of high-order geological geometries (Gómez-Hernández and Wen, 1998; Hansen et al., 2012, 2016; Linde et al., 2017), thus affecting the inversion quality.

Geological realism can be achieved using modeling methods based on training images (TI), which implicitly represent higher-order spatial relationships through gridded data representations. Among these, Multiple-point statistics (MPS) methods (Guardiano and Srivastava, 1993; Mariethoz and Caers, 2014; Strebelle, 2021) are particularly popular in inverse modeling (e.g., González et al., 2008; Melnikova et al., 2015; Høyer et al., 2017; Le Coz et al., 2017), but can be limited by high computational costs and poor posterior exploration due to the large number of subsurface parameters to be optimized. More recently, deep-learning-based generative adversarial networks (GAN; Goodfellow et al., 2014) were demonstrated to be particularly suitable for facies modeling (e.g., Dupont et al., 2018; Laloy et al., 2018; Mosser et al., 2018; Chan and Elsheikh, 2019; Azevedo et al., 2020b; Feng et al., 2022; Miele and Azevedo, 2024). These networks learn to map a low-dimensional latent distribution into the targeted prior distribution of facies patterns represented in a TI. This property makes GANs suitable for the reduction of Bayesian inference computational costs (e.g., Laloy et al., 2017; Chan and Elsheikh, 2019; Levy et al., 2023; Miele et al., 2024). However, their nonlinear transformations negatively affects a full parameters exploration (Laloy et al., 2019; Lopez-Alvis et al., 2021), and multivariate inversion case studies are severely limited by the need for very large networks to represent all the properties of interest (e.g., Wang et al., 2022; Miele et al., 2024). On the other hand, recent designs of GANs employing spatially adaptive denormalization (SPADE-GAN; Park et al., 2019) allow for the local conditioning of image realizations by means of conditioning masks. Abdellatif et al. (2025) demonstrated that SPADE-GANs can be effectively trained for conditional facies modeling, using facies proportion maps as conditioning data. Specifically, at each location within the modeled area, the generated facies class realizations occur with frequencies that reflect the conditioning proportion map. As a result, the proportion map functions as a spatial probability map for each facies class.

In this work, we address probabilistic seismic inversion of facies and rock properties in multivariate nonlinear settings. We propose an iterative inversion method that integrates both SPADE-GAN for facies modeling and geostatistical co-simulation for facies-dependent continuous properties, leveraging both methods' local conditioning capabilities. The method offers a flexible framework for the joint prediction of subsurface properties of different domains, sharing the basic principles of the method proposed by Azevedo et al. (2020a). At each iteration, a pre-trained SPADE-GAN is used to generate a set of facies realizations conditioned on a probability map. The correlated facies-dependent rock properties are then modelled using stochastic sequential co-simulations with multi-local distribution functions (Nunes et al., 2017). The realizations are used to compute the corresponding synthetic seismic data and compared on a trace-by-trace basis against the observed data. At each iteration, model updates maximize a seismic data similarity measure: for each location, the set of properties providing the best fit to the observed data is used to build a new probability model at the following iteration. In the methodology proposed herein, we propose the use of a data similarity measure accounting for data errors under Gaussian model assumptions.

The method is validated on both synthetic and real case studies, for the prediction of facies, porosity, and acoustic impedance from full-stack seismic data. In the synthetic examples, we account for cases where the prior assumptions on facies distributions (i.e., TI and trained SPADE-GAN) are biased compared to the true, target one. The real-case study accounts for a 2-D seismic section from the Norne dataset (North Sea), using locally observed well data as either conditioning or control (blind test), in two separate application examples. These applications demonstrate the algorithm's ability to accurately predict multivariate distributions of categorical and continuous properties conditioned on nonlinear geophysical data. The results obtained for the synthetic cases highlight that the network can find well-fitting solutions even when the prior is biased. The applications to the real case show how the SPADE-GAN generates accurate solutions, which can be significantly improved when conditioning well data are present. Nonetheless, we also note inversion accuracies due to possible network's modeling errors. The proposed method can be adapted to different grid sizes and numbers of subsurface properties with minimal changes to the SPADE-GAN architecture.

## 2. Methodology

The proposed seismic inversion algorithm optimizes ensembles of stochastic subsurface models to predict facies (e.g., litho-fluid or geological) and associated continuous properties. The algorithm builds upon the stochastic perturbation optimization (Azevedo et al., 2020a), mainly integrating a pre-trained SPADE-GAN for facies modeling. In this work, we apply the algorithm for the prediction of facies ($\mathbf{f}$), porosity ($\mathbf{\phi}$) and acoustic impedance ($\mathbf{I_P}$) distributions form full-stack seismic data. Alternative adaptations to different rock-physics modeling (Avseth, 2010; Mavko et al., 2020) and to a varying number of subsurface properties using seismic partial-stacks are also possible, following Azevedo et al. (2020a) and Miele et al. (2023).

A schematic representation of the proposed inversion algorithm, SPADE-GANInv, is shown in Figure 1. In the first step, the SPADE-GAN is trained to model locally conditioned facies distributions based on a prior TI. We use the trained model to generate $N$ facies realizations conditioned on a prior probability map. Facies-dependent, co-located $\mathbf{\phi}$ and $\mathbf{I_P}$ distributions are generated sequentially, using direct sequential co-simulations (co-DSS; Soares, 2001; Horta and Soares, 2010) honoring multi-local prior data distribution and variogram models defined for each facies (Nunes et al., 2017). The $\mathbf{I_P}$ realizations are used to compute the corresponding synthetic full-stack seismic data ($\mathbf{d_{syn}}$), which is trace-by-trace compared to the observed one ($\mathbf{d_{obs}}$). The parameters providing the highest fit to $\mathbf{d_{obs}}$ are selected from the realizations' ensemble to perturb the facies probability map and the geostatistical models; a new set of realizations is then generated at the next iteration. In the following, we first describe the SPADE-GAN and how its training in our application. Then, we describe each SPADE-GANInv step in detail.

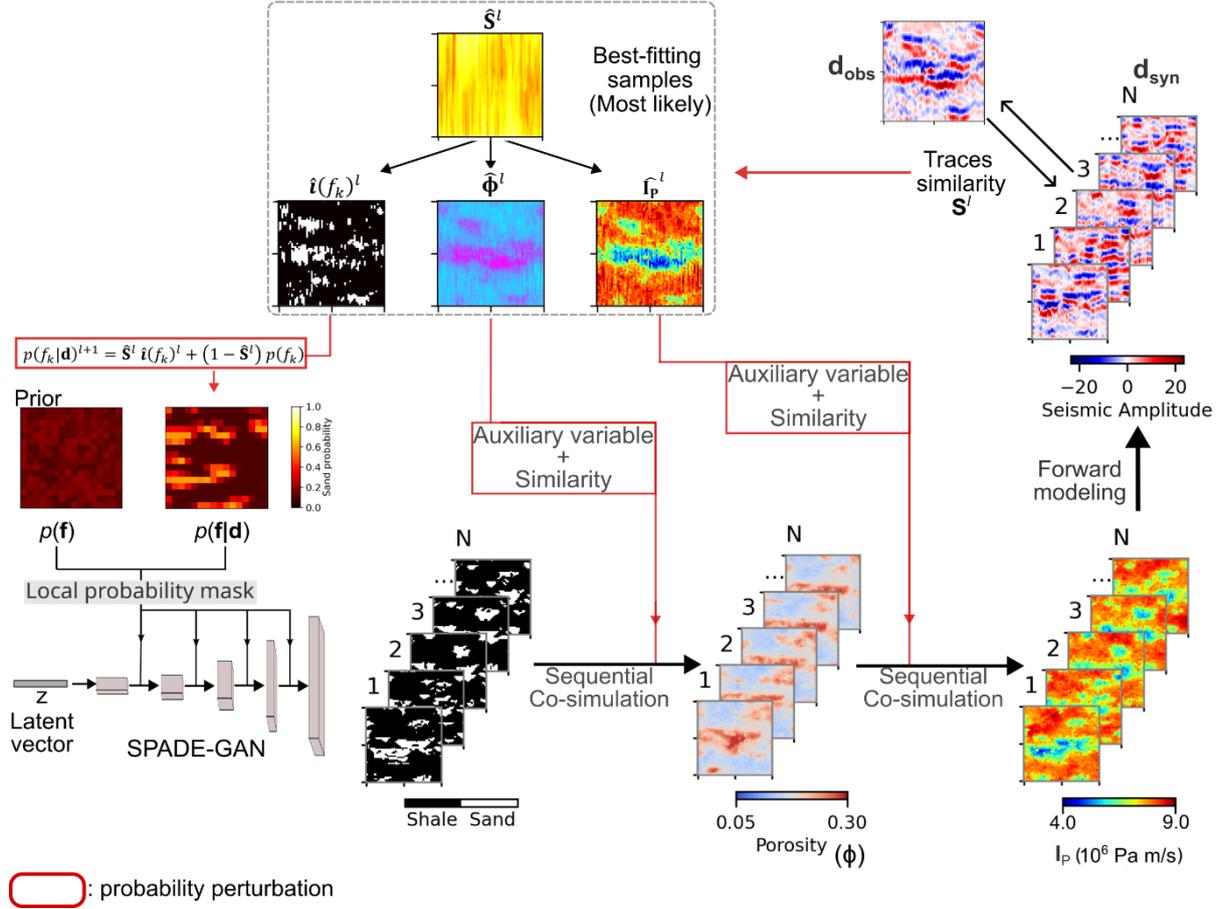

Figure 1: Proposed stochastic seismic inversion algorithm using the pre-trained SPADE-GAN, designed for full-stack seismic data inversion for simultaneous **f**, **ϕ** and **I_P** predictions.

### 2.1. SPADE-GAN

Generative adversarial networks (Goodfellow et al., 2014) are deep generative networks learning a mapping between a parameterized low-dimensional latent space and that of the desired spatial variable distribution, as described in the TI. In its general architecture, the network consists of two components, a generator (*G*) and a discriminator (*D*): *G* is designed to transform a Gaussian low-dimensional latent vector (**z**) into the desired spatial distribution (e.g., facies realizations); *D* encodes the data (both generated and TI samples) into classes or probability values. These two networks are trained in an adversarial framework following the objective function

$$\min_{G(\cdot)} \max_{D(\cdot)} \mathbb{E}[\log D(\mathbf{X})] + \mathbb{E}\left[\log\left(1 - D(G(\mathbf{z}))\right)\right], \tag{1}$$

where **X** is the data from TI and *G*(**z**) are the generated samples. Following Equation 1, *D* is trained to discriminate between generated and TI samples, associating the lowest probability with the generated samples, while *G* is trained to generate samples honoring the TI features, so that the corresponding *D* probability value is maximized. Both *G* and *D* architectures typically involve stacked convolutional layers to capture spatial features at different scales (e.g., LeCun et al., 2015). The intermediate feature maps are normalized using batch normalization (Ioffe and Szegedy, 2015) to stabilize and accelerate training. In the generator, batch normalization is often followed by de-normalization layers or learned affine transformations to restore feature statistics before output generation. The normalized or de-normalized activations are then passed through nonlinear activation functions (e.g., ReLU, LeakyReLU, or Tanh) to introduce nonlinearity and enhance representational capacity.

Spatially-adaptive de-normalization (SPADE; Park et al., 2019) allows locally-conditioned generative modeling with GANs (hence, SPADE-GAN), based on a given conditioning mask (**M**). At each layer of the generator ($l$), output channel ($c$), and location (i.e., pixel or cell – $u$), the SPADE embeds the spatial features of **M** into two additional parameters $\gamma_{l,c,u}(\mathbf{M})$ and $\beta_{l,c,u}(\mathbf{M})$. Therefore, after convolution and batch normalization, the data is rescaled following (Park et al., 2019)

$$\hat{h}_{l,c,u} = \gamma_{l,c,u}(\mathbf{M}) \frac{h_{l,c,u} - \mu_{l,c}}{\sigma_{l,c}} + \beta_{l,c,u}(\mathbf{M}), \qquad \mathbf{M} \in \mathbb{R}^d, \tag{2}$$

where $\mu_{l,c}$ and $\sigma_{l,c}$ are respectively the mean and standard deviation of the batch normalization, and $h_{l,c,u}$ is the output of the convolutional layer. The discriminator receives the conditioning mask **M** as part of its input, concatenated with the image. Training a SPADE-GAN requires a TI having image examples and the associated conditioning mask **M**, and the network learns a conditional distribution. Equation 1 is finally re-written as

$$\min_{G(\cdot)} \max_{D(\cdot)} \mathbb{E}\left[\log D(\mathbf{X}|\mathbf{M})\right] + \mathbb{E}\left[\log\left(1 - D\big(G(\mathbf{z}|\mathbf{M})\big)\right)\right]. \tag{3}$$

### 2.1.1. Network implementation

We adopt the same implementation as Abdellatif et al. (2023) for 2-D image generation. The backbone of the network's layers is the ResNet block (He et al., 2016); Figure 2 shows a schematic representation of the ones in this work for $G$ and $D$. Depending on the specific application, we use a different number of layers. The self-attention mechanism (Vaswani et al., 2017) is further used at an intermediate layer for both $G$ and $D$ (e.g., at resolution $32 \times 32$ for data dimension of $64 \times 64$). The output layer of $G$ is a convolution operator followed by hyperbolic tangent activation function (Tanh). The output of $D$ is produced using a dense linear layer.

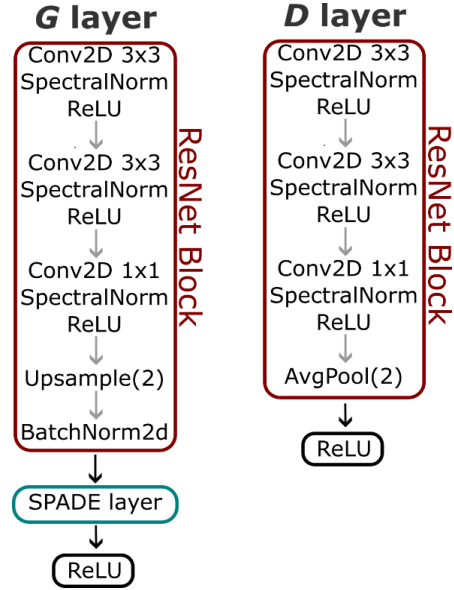

Figure 2: Hidden layers' architecture of the SPADE-GAN used for this work. *Conv2D*: 2D convolutional operation and corresponding kernel size; *SpectralNorm:* spectral normalization; *ReLU:* rectified linear unit function; *AvgPool(2)* Average downscaling with kernel size 2; *Upsample(2)*: upscaling operations (nearest neighbor) with kernel size 2.

### 2.1.2. Facies modeling and network training

A facies class ($f_k$) is described as a discrete variable for each location $u$ as (e.g., Caers and Hoffman, 2006):

$$i(u, f_k) = \begin{cases} 1, & \text{if } f_k \text{ occurs at } u \\ 0, & \text{otherwise} \end{cases}. \tag{4}$$

When more than two facies classes are considered, each $f_k$ can be individually represented by a SPADE-GAN output channel (i.e., using *one-hot encoding*). The spatial distribution of the facies classes is represented by means of gridded data in the TI, derived, for example, from conceptual geological models or through MPS simulation (e.g., Dupont et al., 2018; Laloy et al., 2018; Feng et al., 2022; Miele and Azevedo, 2024; Abdellatif et al., 2025). Additionally, the training data for SPADE-GAN integrates a conditioning mask **M** to each facies training sample. Abdellatif et al. (2025) show that if **M** is a map of facies proportions in a given area, the network parameters $\gamma_{i,c,u}(\mathbf{M})$ and $\beta_{i,c,u}(\mathbf{M})$ can be trained so that $G(\mathbf{z}|\mathbf{M})$ honors the local conditioning proportion, hence $\frac{1}{N}\sum_n^N i(u, f_k)_n \approx \mathbf{M}_{f_k}$. In SPADE-GANInv we consider **M** as a probability map of occurrence of a given $f_k$, with no changes to this mathematical framework. A trained $G$ can be treated as a prior model of occurrence of facies distributions $p(f_k|\mathbf{M}_{f_k})$, when $\mathbf{M}_{f_k}$ is the prior probability of occurrence.

To each facies sample in the TI, we compute the relative probability map **M** by counting the frequency of occurrence of $f_k$ in regions defined by a regular grid. Figure 3a provides an example of a TI with probability maps of resolution $4 \times 4$ calculated on facies samples of resolution $64 \times 64$; the grey squares highlight different facies distributions for the same $\mathbf{M}_{f_k}$.

In our application examples, the SPADE-GAN is trained using hinge loss (Lim and Ye, 2017). The network is optimized using Adam (Kingma and Ba, 2017) with parameters $\beta_1 = 0.5$ and $\beta_2 = 0.999$, a learning rate of $2 \times 10^{-3}$ for both $G$ and $D$ with a fixed step decay of 90% every 100 epochs. The total number of training epochs mostly depended on the specific application; the training is stopped when we detect stable adversarial loss values along 100 epochs, and the fit between $G(\mathbf{z}|\mathbf{M})$ and the conditioning facies map reaches a stable maximum value on the validation data.

### 2.2. SPADE-GANInv

#### 2.2.1. Facies simulations

At each iteration ($l$) of the proposed inversion algorithm, a set of $N$ facies realizations are generated using the pre-trained SPADE-GAN, conditioned on the probability map $\mathbf{M}^l$, hence $G(\mathbf{z}|\mathbf{M}^l) \approx p(\mathbf{f}|\mathbf{M}^l)$ and $\mathbf{f}^{l,r} \sim G(\mathbf{z}^{l,r}|\mathbf{M}^l)$, $r = [1, \dots, N]$. At the first iteration, $\mathbf{M}^l$ is the prior facies distribution. In cases of considerable lack of knowledge, $\mathbf{M}^l$ can be set as uniform for the whole study area. The generated realizations reproduce the facies spatial patterns statistics, and the conditioning probability used as input.

#### 2.2.2. Continuous properties simulations

Each $\mathbf{f}^{l,r}$ are used to locally condition the sequential geostatistical simulation of $\boldsymbol{\phi}$ and $\mathbf{I_P}$ using co-DSS with multi-local distribution functions (Nunes et al., 2017). This method allows generating realizations of subsurface properties honoring marginal and joint distribution of the model variables, and their spatial distribution model (i.e., variogram model), as defined from the available well-data, for each facies class. In such multivariate scenario, we first simulate a porosity realization $\boldsymbol{\phi}^{l,r}$, conditioned on $\mathbf{f}^{l,r}$, as $\boldsymbol{\phi}^{l,r} \sim F_\phi(\boldsymbol{\phi}|\mathbf{f}^{l,r})$; then, we simulate acoustic impedance as $\mathbf{I_P}^{l,r} \sim F_{IP}(\mathbf{I_P}|\mathbf{f}^{l,r}, \boldsymbol{\phi}^{l,r})$. $F_\phi$ and $F_{IP}$ are the properties-specific conditional distributions. The mapping from petrophysical to elastic domain (e.g., from $\boldsymbol{\phi}^{l,r}$ to $\mathbf{I_P}^{l,r}$) can be also performed using rock physics modeling (e.g., Azevedo et al., 2019, 2020a; Mavko et al., 2020; Miele et al., 2023).

#### 2.2.3. Geophysical forward model

The geostatistical realizations of elastic properties are used for the calculation of $N$ synthetic seismic data ($\mathbf{d}_{\mathbf{syn}}^{l,r}$). In full-stack seismic data inversion, we use the $\mathbf{I_P}$ realizations to compute the corresponding reflectivity sequences for normal incidence, given by the contrasts of impedance between the $i$-th trace sample and the following

$$rc_i = \frac{I_{P\,i+1} - I_{P\,i}}{I_{P\,i+1} + I_{P\,i}}. \tag{5}$$

Then, the reflection coefficients are convolved with a wavelet representative of the recorded seismic data to generate the synthetic seismic.

*2.2.4. Optimization*

The objective function aims at the maximization of the seismic data traces between $\mathbf{d_{syn}}$ and the observed data $\mathbf{d_{obs}}$. In this work we consider an objective function accounting for uncorrelated Gaussian data error, defining a data similarity ($S$) function

$$S = \exp\left(-\frac{1}{2M}\sum_{i=1}^{M}\frac{d_{obs,i}^2 - d_{syn,i}^2}{\sigma_i^2}\right), \tag{6}$$

where $M$ is the number of data samples in the traces (or traces portion) considered and $\sigma_i^2$ is the data error variance. Equation 6 is a monotonic transformation of the negative Gaussian log-likelihood, providing a similarity value bounded in the range [0,1] ($S = 1$ when $\mathbf{d_{syn}} = \mathbf{d_{obs}}$). In Equation 6 we remove the indexes $l$ and $r$ for better readability.

For each realization, local values of $S$ are evaluated on seismic traces portions of random length $M$, ranging from one quarter to 1.5 times the source wavelet length. To account for variability introduced by the random section size, this procedure is repeated multiple times and averaged at each location. Finally, the subsurface properties realizations providing the largest local similarity ($\hat{\mathbf{S}}^l$) are stored ($\hat{\imath}(f_k)^l, \hat{\boldsymbol{\phi}}^l, \widehat{\mathbf{I_P}}^l$) to perturb the probabilities of subsurface models at the next iteration.

The facies conditioning probability is updated using (Caers and Hoffman, 2006)

$$p(f_k|\mathbf{d})^{l+1} = \hat{\mathbf{S}}^l\,\hat{\imath}(f_k)^l + \left(1 - \hat{\mathbf{S}}^l\right)p(f_k). \tag{7}$$

Therefore, we use the probability map resulting from Equation 7 for the generation of a new set of $N$ facies realizations. The model perturbation of the geostatistical co-simulation of continuous properties is performed using $\hat{\boldsymbol{\phi}}^l$ and $\widehat{\mathbf{I_P}}^l$ as auxiliary variables and $\hat{\mathbf{S}}^l$ as their corresponding local correlation coefficient, so that $\boldsymbol{\phi}^{l+1,r} \sim F_\phi(\boldsymbol{\phi}|\mathbf{f}^{l+1,r}, \hat{\mathbf{S}}^l)$ and $\mathbf{I_P}^{l+1,r} \sim F_{IP}(\mathbf{I_P}|\mathbf{f}^{l+1,r}, \boldsymbol{\phi}^{l+1,r}, \hat{\mathbf{S}}^l)$ – we refer to Azevedo et al., (2020a) and Azevedo and Soares (2017) for details on the geostatistical models' perturbation.

Finally, following Equation 6, $S = 1$ when $\mathbf{d_{syn}}$ overfits to the data noise. For this reason, the optimization is regularized using the weighted root mean squared error (WRMSE)

$$\text{WRMSE} = \sqrt{\frac{1}{M}\sum_{i=1}^{M}\frac{d_{obs,i}^2 - d_{syn,i}^2}{\sigma_i^2}}. \tag{8}$$

That is, when the WRMSE < 1 the prediction error is lower than the standard distribution of the data error, and we stop updating the probabilities for that given location. We repeat the optimization until a tolerance fit value is met (e.g., all the traces have WRMSE=1), or when a maximum number of pre-defined iterations is reached.

The resulting subsurface realizations in this iterative procedure honour the prior geological knowledge and the conditioning data (i.e., training image and borehole data, respectively), under the geophysical forward model and data error assumptions.

## 3. Synthetic Application examples

For the synthetic application examples, we consider a geological scenario representing sand bodies deposited in a shale background. The morphology of the channels is represented in the TI using 3000 2-D MPS facies realizations (Figure 3a) of size $64 \times 64$ cells. Each realization is associated to a probability distribution map of resolution $4 \times 4$ pixels (Figure 3a), obtained by calculating the occurrence of sand channels in areas of $16 \times 16$ cells each facies sample. The prior TI sand volume fraction is $0.21 \pm 0.05$, and the sand bodies are uniformly distributed over the investigated area (Figure 3b): this defines the prior $\mathbf{M}^l$ used in inversion.

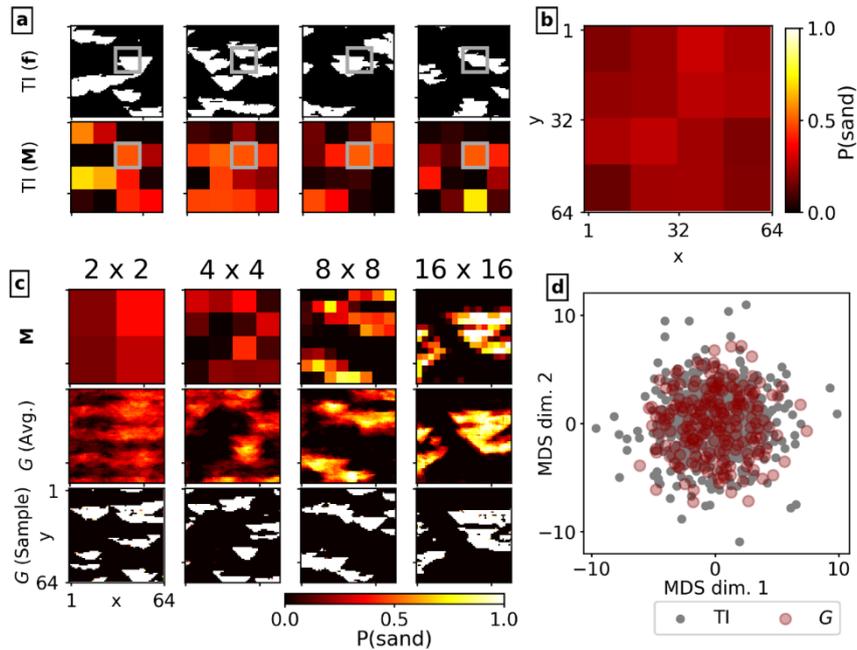

Figure 3: a) Training data facies examples, TI ($\mathbf{f}$), and corresponding probability mask, TI ($\mathbf{M}$) – the gray squares highlight different TI ($\mathbf{f}$) samples honoring the local $p(sand) = 0.5$ in TI ($\mathbf{M}$); b) Example of prior low-resolution conditioning mask; c) examples of generated realizations' mean, $G$ (Avg.), and samples, $G$ (Samples), given $\mathbf{M}$ of different resolutions, using the trained SPADE-GAN for the synthetic case; d) MDS space computed with pairwise Hausdorff distances using TI samples (gray) and generated realizations of the SPADE-GAN (red).

### 3.1. Prior data modeling and SPADE GAN training

The SPADE-GAN accounts for five layers in $G$ and four layers in $D$ (Figure 2). The encoded latent vector $\mathbf{z}$ has size $512$. We trained the network for 1000 epochs, reaching training stability after 600 epochs. The total training time on a Linux machine, using a single GPU NVIDIA® GeForce GTX TITAN X, was approximately 25h.

The modeling accuracy is evaluated on the SPADE-GAN ability to reproduce the TI spatial patterns. We first sampled random $\mathbf{M}^l$ from a multivariate normal distribution defined on the prior volume fraction $\mathcal{N}(\mu = 0.21, \sigma^2 = 0.05^2 I_d)$, with different resolutions (e.g., Figure 3c) for prior facies generation. The SPADE-GAN realizations reproduced the training facies spatial patterns and honor the local conditioning probability (Figure 3c) with an average error between the conditioning probability **M** and the corresponding generated facies occurrence $G(\mathbf{z}|\mathbf{M})$ of 6%. These observations confirm that modeling is possible using conditioning maps of different resolutions from training (Abdellatif et al. 2025). The pairwise Hausdorff distances between generated and training images is projected as a 2-D distribution through multi-dimensional scaling (MDS; Cox and Cox, 2008) in Figure 3d. The overlap between TI (gray circles) and SPADE-GAN samples (red circles) indicates good fitting of facies morphologies and similar distributions variability.

### 3.2. Seismic data inversion

We consider four different synthetic scenarios for the validation of SPADE-GANInv (Figure 4). The facies distributions (Figure 4a) for Target 1, 2, and 3 are test data extracted from the TI; Target 4 uses an arbitrary facies distribution that is particularly biased from the prior, with a sand volume fraction of **0.02**. For these samples, Hausdorff distances from the TI are plotted in in a 2-D MDS space in Figure 4b.

The corresponding $\boldsymbol{\phi}$ and $\mathbf{I_P}$ have marginal and joint distributions sampled from multiGaussian distributions (Figure 4c). Specifically, $\boldsymbol{\phi}_{shale} = 0.17 \pm 0.02$, $\mathbf{I_P}_{shale} = (8.0 \pm 0.4)$ Pa s/m, $\boldsymbol{\phi}_{sand} = 0.25 \pm 0.02$, and $\mathbf{I_P}_{sand} = (6.5 \pm 0.3)$ Pa s/m; their linear inverse correlation is affected by bi-variate Gaussian noise, representing uncertainty on the rock physics model. For each facies class we use a separate variogram model to describe $\boldsymbol{\phi}$ and $\mathbf{I_P}$ spatial patterns. For shale we use a spherical variogram with ranges of 30 and 10 cells along the horizontal and vertical directions, respectively. For sands we define a spherical, isotropic variogram model with 40 cells range. Both consider no nugget effect. During the inversion process, we adopt the same parameters for geostatistical co-simulation. Each $\mathbf{d_{obs}}$ is computed from $\mathbf{I_P}$ distributions using Equation 5 and a Ricker wavelet with central frequency of 25 Hz and length of 60ms. We assume data noise of magnitude $\sigma_{\mathbf{d}} = 0.5 \pm 0.05 * |\mathbf{d_{obs}}|$.

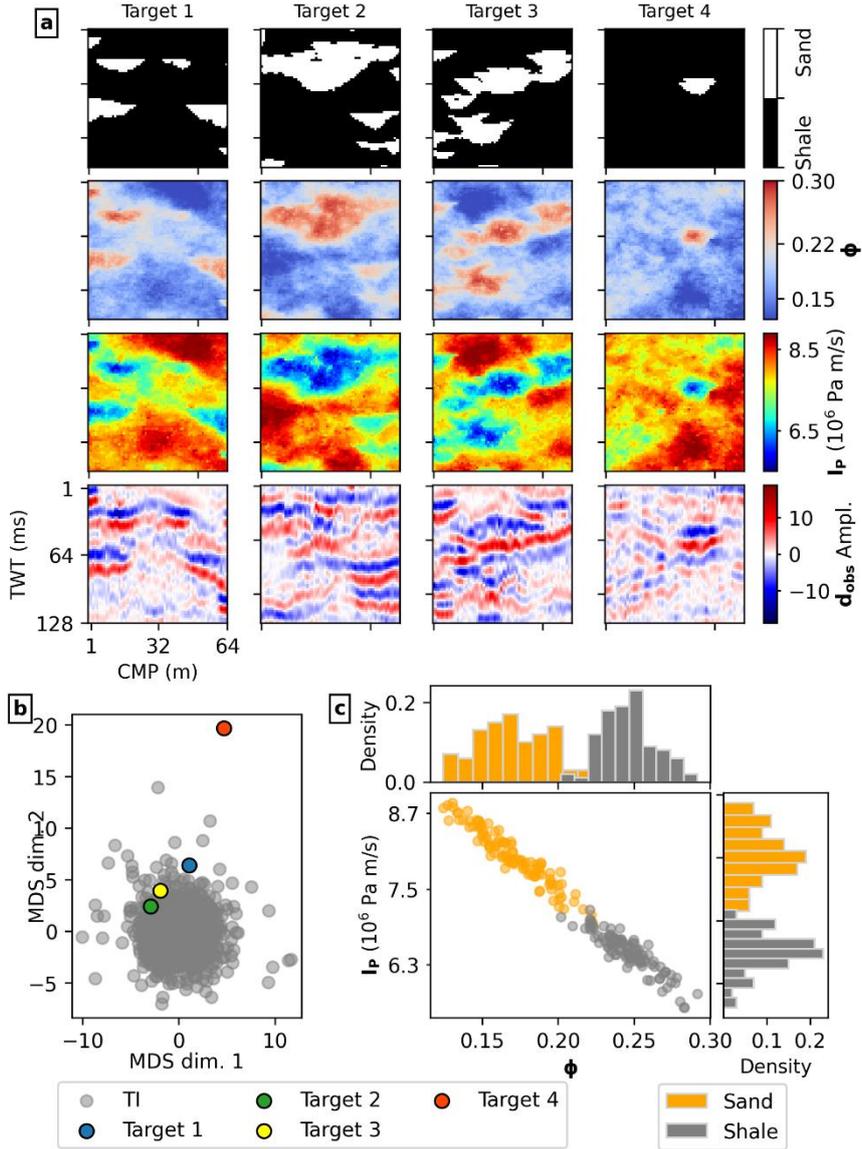

Figure 4: Synthetic case application examples; a) **f**, **ϕ**, **I_P** and corresponding **d_obs**, for the five cases considered ("Target"); b) MDS space computed with pairwise Hausdorff distances for TI samples and target facies samples; c) prior marginal and joint distribution of **ϕ** and **I_P** considered for each target.

We ran the inversion for 300 iterations, generating 300 realizations each. In our experiments, took approximately 2.5h per test case running entirely on a CPU Intel® Core™ Ultra 7 165H. An example of the iterative update of the conditioning probability map ($\mathbf{M^l}$) is given in Figure 5 for Target 1. The prior map (*Iter. 1* in Figure 5) is randomly sampled the prior $\mathcal{N}(\mu = 0.21, \sigma^2 = 0.05^2 I_d)$. The updated maps (*Iter.* >1) progressively approach the true distribution, showing few residuals and vertical patterns due to the trace-based similarity. The SPADE-GAN generated facies realizations consistent with the conditioning probability and the learned prior features. The influence of the latter likely explains the local mismatches observed between the conditioning probabilities and the generated realizations (as further observed and discussed below).

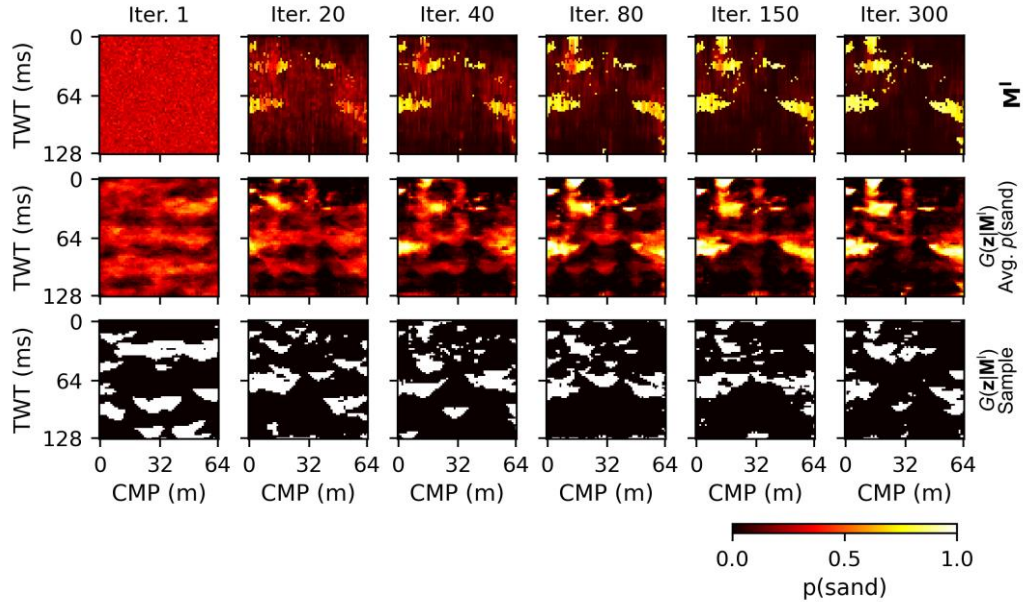

Figure 5: Example of the iterative update of the conditioning facies probability map ($\mathbf{M}^l$) for SPADE and the corresponding generated distributions ($G(\mathbf{z}|\mathbf{M}^l)$) per inversion iteration (*Iter.*); *Avg.*: point-wise average from 30 realizations; *Sample*: single realization.

The inversion results are summarized in Figure 6, Figure 7 and Table 1; all metrics and figures are based on 300 post-inversion realizations. The SPADE-GANInv converged closely to the target distributions for all properties (Table 1 and Figure 7). The results for Target 4 shows the highest data error ($\text{WRMSE}_\mathbf{d}$ in Table 1), as expected from the mismatch between the true facies and the assumed prior. Nonetheless, the realizations correctly predict the facies classes of each pixel 86% of the time, on average, yielding p(sand) and most-likely facies maps that closely match the target (Figure 6). The average sand volume fraction predicted is, however, overestimated at **0.15**. In MDS space (Hausdorff distances; Figure 7a), these realizations fall within the prior-pattern range and cluster to minimize the distance from the target. Conversely, Target 2 exhibits the lowest facies-prediction accuracy, failing to capture the large upper sand bodies and underestimating their lateral continuity (Figure 6). This mismatch is likely due to the scarcity of such large, continuous sand patterns in the training dataset.

The $\boldsymbol{\phi}$ and $\mathbf{I_P}$ predictions generally honor the conditioning data distributions (Figure 7b) and their average distributions match the target (Figure 6). The lowest $\text{RMSE}_{\boldsymbol{\phi}}$ and $\text{RMSE}_{\mathbf{I_P}}$ values are found for Target 1 and Target 3 cases (Table 1). A comparison of the inversion predictions along a random vertical trace between the test case 1 and 4 is provided in Figures 7c and 7d. For Target 1, 95% of both $\boldsymbol{\phi}$ and $\mathbf{I_P}$ true values are enclosed within the predicted distributions; for Target 4, the data within this range is 85%.

The SPADE-GAN generated facies realizations show a consistent horizontal pattern, common to all case studies (e.g., shown in the *p*(sand) of Figure 5 and Figure 6). Despite the optimized training procedure used, these trends are likely to be network's artifacts, which can have an influence on the inversion results. In this case study, their complex, nonlinear distribution prevented improving the inversion results with appropriate network errors modeling.

Table 1: accuracy and precision metrics of the results from the synthetic case application of the proposed inversion algorithm.

|  | Target 1 | Target 2 | Target 3 | Target 4 |
|---|---|---|---|---|
| $WRMSE_\mathbf{d}$ | 6.1 ± 0.5 | 6.9 ± 0.5 | 5.9 ± 0.4 | 8.0 ± 0.5 |
| $\mathbf{f}$ accuracy | 85% | 78% | 83% | 86% |
| $RMSE_{\boldsymbol{\phi}}$ ($\times 10^{-2}$) | 2.4 ± 0.1 | 2.8 ± 0.2 | 2.6 ± 0.2 | 3.2 ± 0.1 |
| True $\boldsymbol{\phi}$ in $10^{th}$–$90^{th}$ | 70.0% | 67.5% | 61.1% | 69.7% |
| $RMSE_{I_P}$ ($\times 10^5$) | 5.3 ± 0.2 | 5.5 ± 0.4 | 5.1 ± 0.3 | 6.4 ± 0.2 |
| True $\boldsymbol{\phi}$ in $10^{th}$–$90^{th}$ | 72.8% | 69.4% | 63.0% | 71.2% |

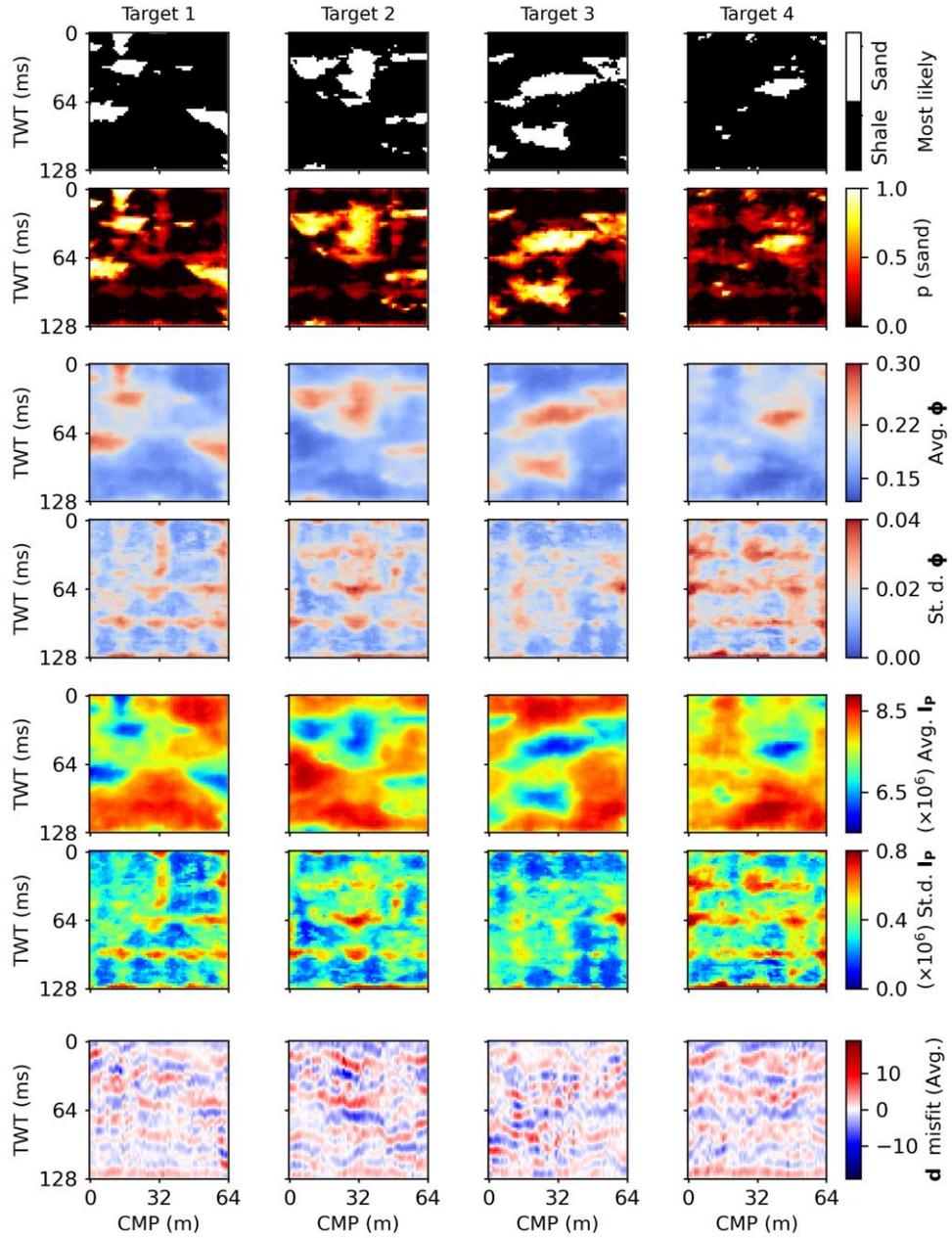

Figure 6: Results obtained for the seismic inversion on synthetic cases; *Avg.*: average; *St.d.*: standard deviation. All the results are computed from the realizations' ensemble generated at the last iteration.

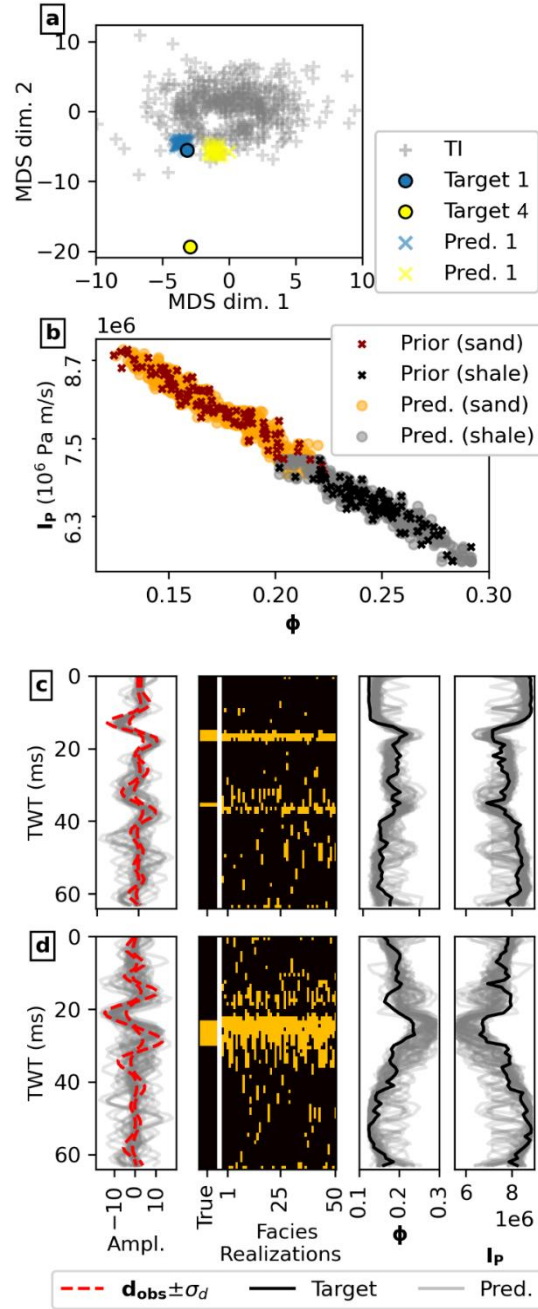

Figure 7: a) MDS projection of Hausdorff distances between prior, targets and predicted facies realizations for the synthetic study cases 1 and 4; b) comparison between the target ("Prior") and the predicted ("Pred.") $\boldsymbol{\phi} - \mathbf{I_P}$ joint distribution; c) and d) synthetic seismic data and predicted $\mathbf{f}$, $\boldsymbol{\phi}$ and $\mathbf{I_P}$ for the targets 1 (c) and 4 (d), for a vertical trace located at $x = 50\,m$.

## 4. Real case application example

We further applied SPADE-GANInv to a real-case study, considering a 2-D seismic section (Figure 8c) acquired in the Norwegian Norne Field (e.g., Rwechungura et al., 2010). The area extends for 1.6 km laterally and 254 ms TWT vertically. As each seismic trace is 12.5 m distant from the other and the seismic resolution is 2 ms, we defined an inversion grid of $128 \times 128$ cells. The local geology is summarized as a sequence of predominantly sandy formations intercalated with shaly layers, mostly representing intercalated marine and continental deposits (Rwechungura et al., 2010; Santiago, 2015). The observed seismic trace crosses a well (Figure 8c) from which facies, $\phi$ and $I_P$ logs were collected (Figure 8d). The correlation between $\phi$ and $I_P$ shows significant uncertainty (Figure 8e). To model their spatial distribution, we fit the variogram models using both well log data for their vertical direction, and the seismic data reflections as an approximate proxy for their the horizonal direction (e.g., Tylor-Jones and Azevedo, 2024). Two exponential variogram models were defined for shale and sand: for shales, the modelled ranges are 30 cells (or 375m) and 24 cells (or 48ms) for the horizontal and vertical directions, respectively; for sands, these ranges are 40 cells (or 500m) and 40 cells (or 80ms).

The source wavelet used is a statistical wavelet available from the original dataset; at well location, the synthetic seismic trace obtained using the available $I_P$ log (in red in Figure 8d) has an $RMSE_d = 150$. We partially include this misfit considering the data uncertainty to have magnitude $\sigma_d = 80 \pm 0.1 * |d_{obs}|$. The source wavelet used is a statistical wavelet available from the original dataset; at well location, the synthetic seismic trace obtained using the available $I_P$ log (in red in Figure 8d) has an $RMSE_d = 150$. We partially include this misfit considering the data uncertainty to have magnitude $\sigma_d = 80 \pm 0.1 * |d_{obs}|$.

### 4.1. Prior facies data and SPADE-GAN training

Through previous studies on the same data (Rwechungura et al., 2010; Santiago, 2015; Azevedo et al., 2019; Miele and Azevedo 2024) and given the sand-to-shale ratio from in the well logs, we generated 6000 facies TI through object modeling using Petrel™ software (SLB©). The probability maps have a spatial resolution of $8 \times 8$ cells (Figure 8a). Compared to the synthetic case, we train a larger SPADE-GAN for $128 \times 128$ outputs, using six layers in $G$ and five layers in $D$ (Figure 2), and sampling from a latent vector of size 1024. The total training time was set to 1000 epochs, and stability (maximum performances) were reached after 500 epochs. Using the same computer settings as in the synthetic case training, the process took approximately 75h. As for the synthetic case, we observe a good fit between TI and generated facies patterns (e.g., Figure 8b) with an average error between the conditioning probability $M$ and the corresponding generated facies occurrence $G(z|M)$ of 5%.

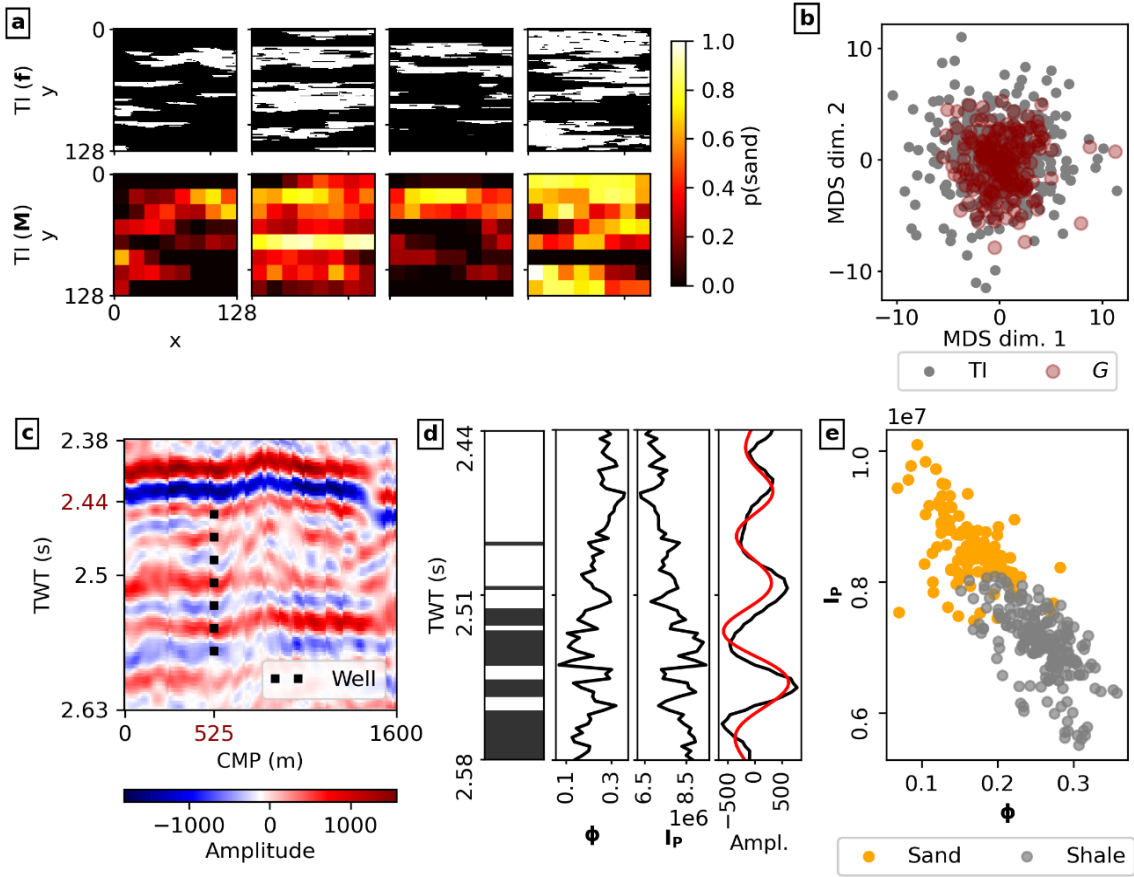

Figure 8: a) examples of the facies realizations from the training dataset for the Norne case; b) MDS space computed with pairwise Hausdorff distances for TI samples and unconditioned samples generated by the trained SPADE-GAN; c) **d_obs** data and well data location; d) Well logs of facies (black: shales; white: sand), **ϕ** and **I_P**, and the co-located **d_obs** (black) and **d_syn** computed from the **I_P** log (red); e) facies-dependent joint distribution between **ϕ** and **I_P** well log data.

### 4.2. Seismic data inversion

In this case we ran two different inversions, one using the well data as conditioning, and one using it for blind testing, referred to as "conditioned" and "unconditioned", respectively. For both, the inversion ran for a total of 300 iterations, each with 150 realizations, on a CPU Intel® Core™ Ultra 7 165H (16core). The inversion algorithm took approximately 3h per test case.

The inversion results are summarized in Figures 9–11, and in Table 2. Both conditioned and unconditioned inversions converged towards solutions with similar **RMSE_d** (Table 2). The optimized facies probability maps (**M**) obtained in each case, and corresponding facies realizations, are shown in Figure 9a. Considering the pair-wise Hausdorff distance between predicted realizations and TI projected on a 2-D MDS space (Figure 9b), the two case studies' results plot into two very close clusters, indicating that for the two cases similar facies patterns where predicted, within prior assumptions.

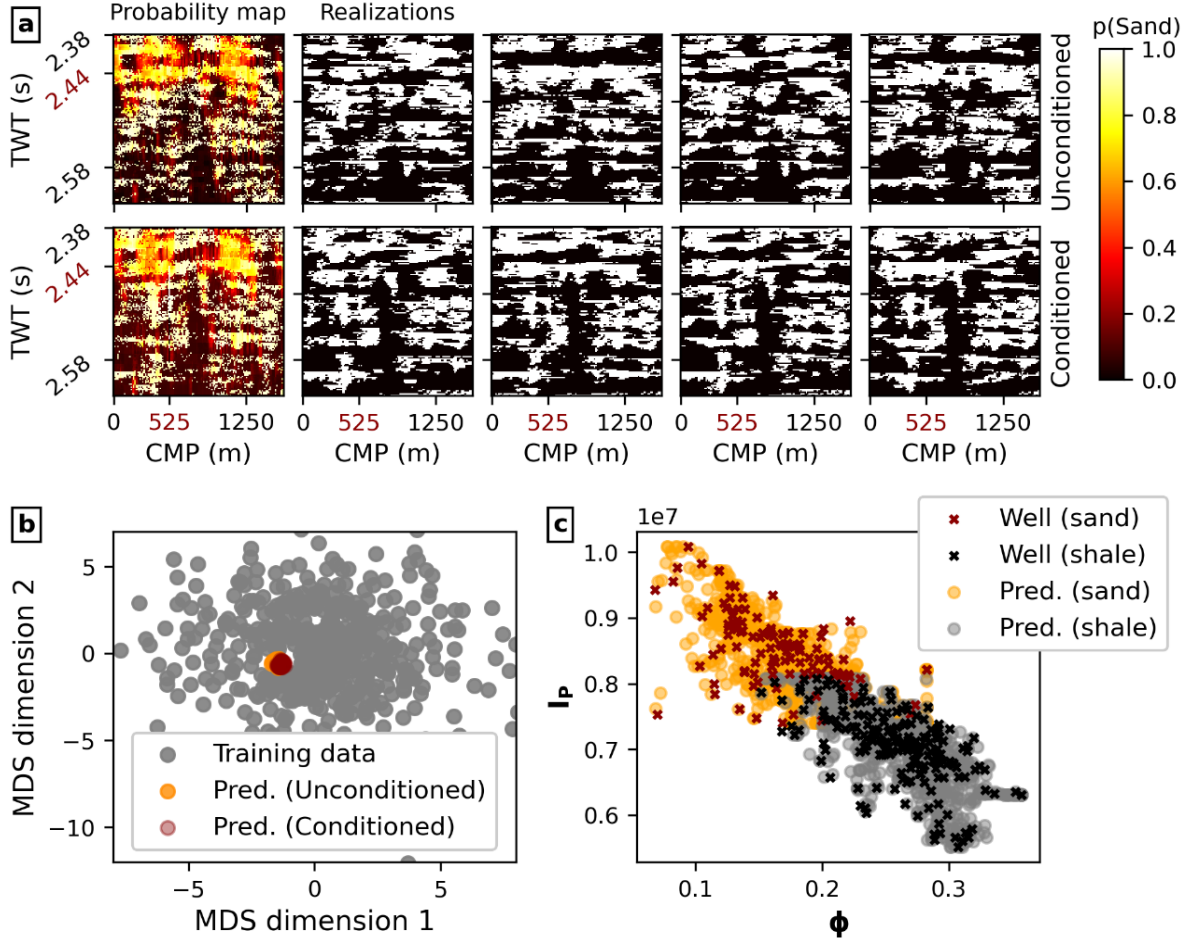

Figure 9: a) Predicted facies probability maps and corresponding realizations generated with the trained SPADE-GAN; b) MDS-projected pair-wise Hausdorff distances between 1000 prior (*Training data*) and 300 predicted facies images from both *Conditioned* and *Unconditioned* cases; c) reproduction of the observed $\boldsymbol{\phi} - \mathbf{I_P}$ joint distribution in the predicted subsurface Co-DSS realizations. Red coordinates in a) indicate the well log head position.

The corresponding $p$(sand) (Figure 10a) shows largest variability located in an upper region (between 2.38 and 2.44 s TWT), roughly corresponding to the location of largest seismic reflections (Figure 8c). Such variability is also reflected in the predicted $\boldsymbol{\phi}$ (Figure 10b), $\mathbf{I_P}$ (Figure 10c) and seismic data error (Figure 10d). The influence of the conditioning well data influences positively the overall accuracy of the predictions (Table 2) and reduces the variability of the results (e.g., *St. dev.* distributions in Figure 10).

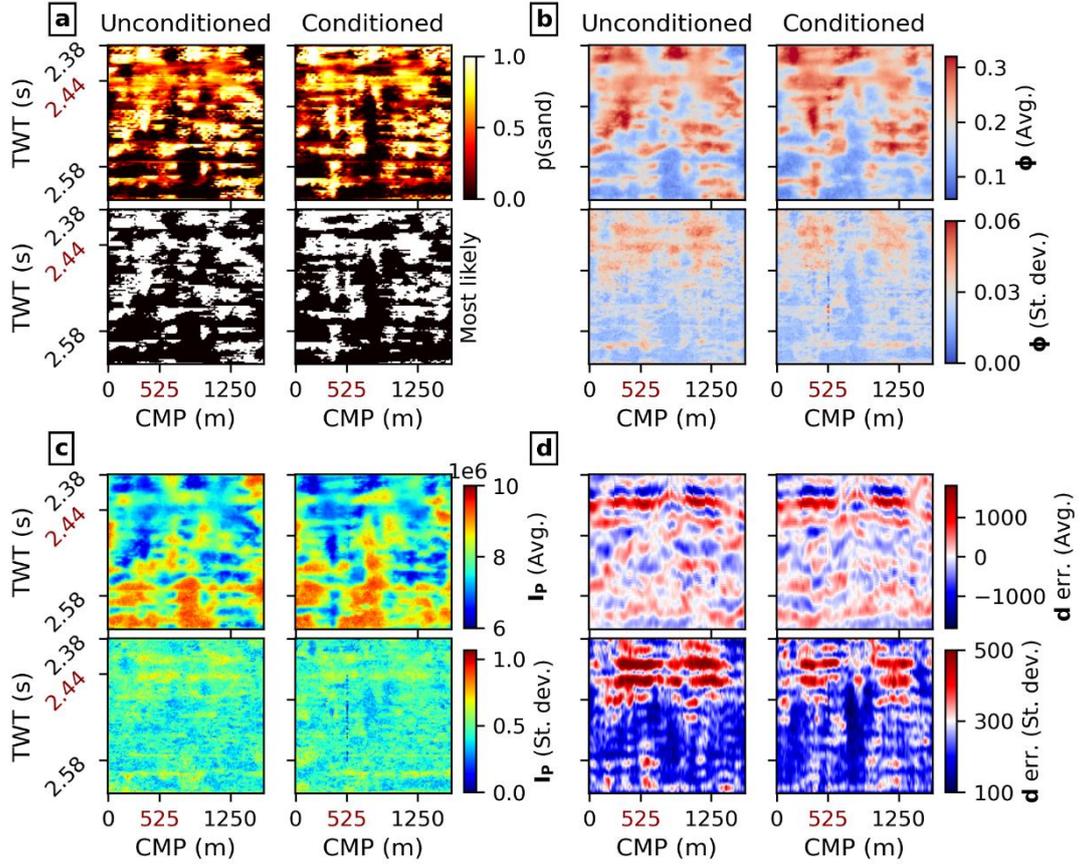

Figure 10: results for the real case inversion applications; a) $p$(sand) and point-wise *most likely* facies; average (*Avg.*) and standard deviation (*St. dev.*) of the predicted b) $\phi$ and c) $I_P$, and computed d) seismic data error. Red coordinates indicate the well log head position. The results are obtained from the realizations' ensemble.

The results obtained at the well location for the unconditioned case (Figure 11a) show how the inversion method retrieved the average trends of the observed data, particularly for **TWT > 2.47 s**. A good percentage of the true well data is represented between the $10^{th}$ and $90^{th}$ percentiles of the predicted $\phi$ and $I_P$ (Table 2 and Figure 11a). As previously observed, the upper section shows larger variability of the predicted properties. Using well data in the conditioning map of the SPADE-GAN (Figure 11b) influences accordingly the generation of facies and the corresponding $p$(sand). However, the generated facies show a mismatch from the conditioning data, clear in the *most likely* distribution. Overall, the predicted $\phi$ and $I_P$ match very well the true data trends (Figure 11b and Table 2). The facies modeling error shown for the conditioned case, affects the modeling of the co-dependent $\phi$ and $I_P$, which deviate from well data values where facies errors are found (Figure 11).

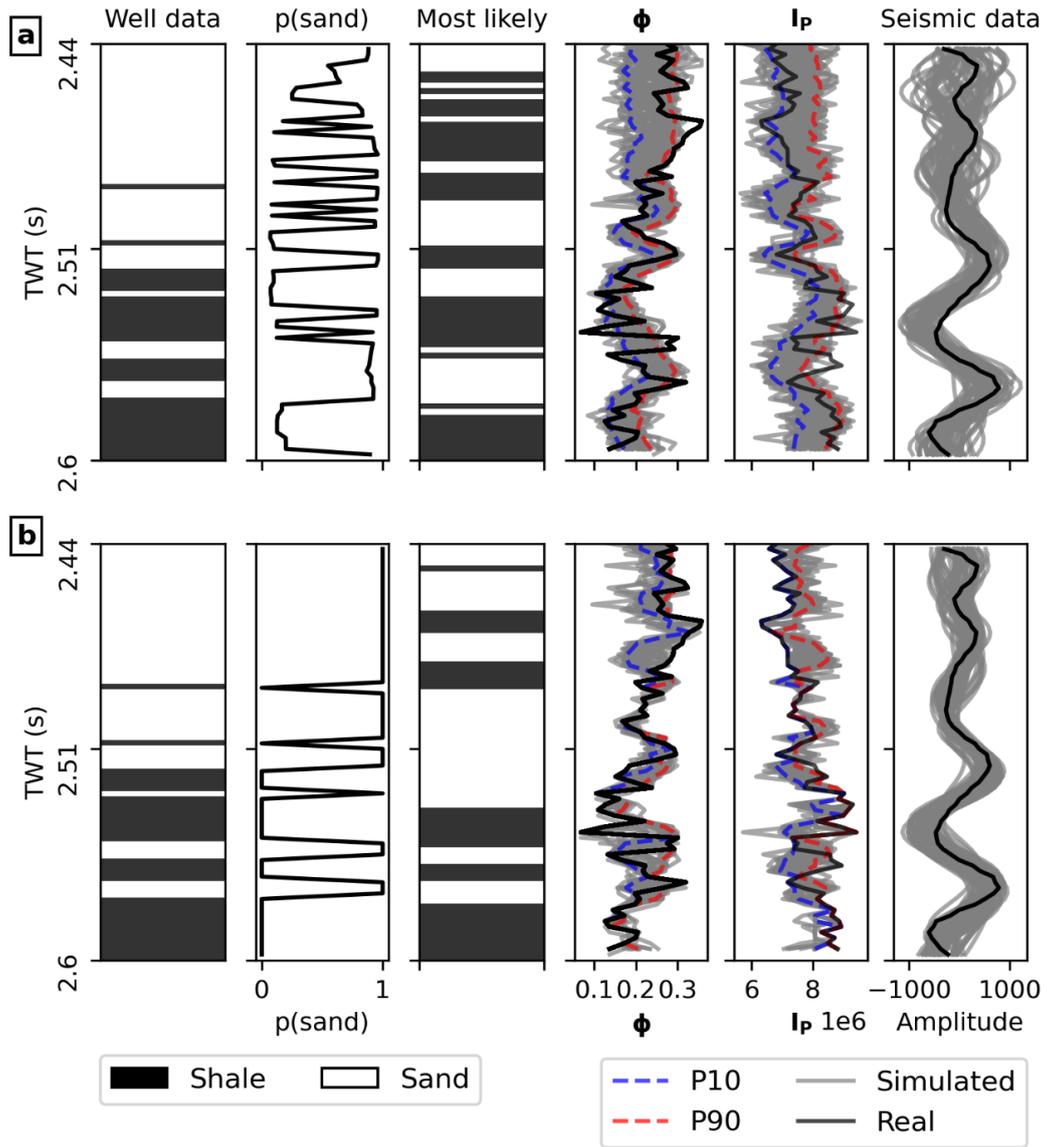

Figure 11: real and predicted properties at the well location for both a) conditioned and b) unconditioned real cases, obtained from the realizations' ensemble at the last iteration. "P10" and "P90" indicate the $10^{th}$ and $90^{th}$ percentiles computed from the realizations' ensembles.

Table 2: accuracy and precision metrics of the results from the real case application of the proposed inversion algorithm.

|  | Unconditioned | Conditioned |
|---|---|---|
| $WRMSE_\mathbf{d}$ | 3.8 ± 0.2 | 3.6 ± 0.2 |
| $\mathbf{f}$ accuracy | 56% | 67.1% |
| $RMSE_{\boldsymbol{\phi}}$ ($\times 10^{-2}$) | 6.1 ± 0.4 | 4.4 ± 0.3 |
| Well $\boldsymbol{\phi}$ in $10^{th}$–$90^{th}$ | 44.6% | 94.5% |
| $RMSE_{\mathbf{I_P}}$ ($\times 10^5$) | 7.9 ± 0.6 | 5.5 ± 0.4 |
| Well $\mathbf{I_P}$ in $10^{th}$–$90^{th}$ | 55.4% | 91.2% |

## 5. Discussion

The proposed seismic inversion method combines a deep-learning method for facies (SPADE-GAN) with geostatistical modeling of continuous properties. Each subsurface property's conditioning probabilities are updated independently using a common objective function: the maximization of a local data similarity measure. This independence allows the joint prediction of multiple correlated subsurface properties, maintaining computational flexibility. Furthermore, by using SPADE-GAN for facies and geostatistical co-simulation for continuous properties, the method avoids the need for very large generative models (e.g., Miele et al., 2024) and is flexibly applicable to multiple properties.

Although the SPADE-GAN was trained on trained on lower-resolution probability maps (Figures 3a and 8a), we can use the higher-resolution, trace-by-trace seismic data similarity to generate new facies probability maps (Figure 5). Analogously, local conditioning on well log data is also possible (Figures 10 and 11b), though the SPADE-GAN does not fully reproduce the exact facies at well locations (e.g., facies accuracy in Table 2 and most-likely facies in Figure 11b). We attribute these modeling errors to two main factors: the influence of facies' probabilities from neighboring locations and the insufficient sensitivity of the SPADE-GAN to sharp lateral changes at pixel scale. These errors introduce undesired variance in the co-located continuous properties at well locations (Figure 11b).

In cases of biased prior assumptions, such as Target 4 in the synthetic examples, the inversion method performed particularly well, retrieving accurate solutions. The SPADE-GAN generated realizations that honored the prior facies features (Figure 7a) while closely approximating the target model (Figure 6) with good accuracy (Figure 7d, Table 1).

In the real-data applications, the method produced realistic results consistent with both well logs and $\mathbf{d_{obs}}$ (Figure 10, Table 2), demonstrating scalability to larger and more complex settings. In areas where the algorithm did not converge (e.g., the upper part of the study area), the results tended to revert to the prior distribution. The lack of convergence in this region is likely due to limited well log data and/or incorrect assumptions in the convolutional model (e.g., the estimated source wavelet), also shown by the mismatch between seismic data at well location in Figure 8d.

The modeling errors of the SPADE-GAN are mainly due to mismatches between the conditioning *p*(sand) and the generated realizations. In the synthetic case, these errors appear as undesired fixed patterns in the realizations (Figure 6), while in the real case, the generated facies models do not fully honor the conditioning well data (Figure 11b). Such effects constrain or bias subsurface parameter exploration and likely explain the low prediction accuracy for Target 2. These modeling errors are difficult to predict or correct within the inversion algorithm, as they can be irregularly distributed and exhibit highly nonlinear correlations with the geophysical domain. This limitation could be mitigated with a more comprehensive training dataset, for instance by incorporating probability maps at multiple resolutions and increasing the number of training examples.

## 6. Conclusions

In this work, we introduce a seismic inversion framework that integrates SPADE-GANs for facies modeling with geostatistical simulation of continuous rock properties. By iteratively updating both categorical and continuous subsurface variables through comparison of synthetic and observed seismic data, the method effectively handles nonlinear multivariate inversion problems. The method is successfully demonstrated for the prediction of facies, $\phi$ and $I_P$ from full-stack seismic data, on both synthetic and real scenarios of different complexities, including a case of biased prior assumptions. The results show that SPADE-GAN can predict the target subsurface properties distributions with good accuracy, while preserving prior geological realism. The inversion outcomes are further improved when conditioning on local well data. We further observe that limitations to this method can arise from the SPADE-GAN training performances, mostly affecting the honoring of the conditioning probability and the introduction of fixed facies patterns. These issues highlight the importance of carefully designing training datasets and improving sensitivity to high-resolution probability variations. Overall, the proposed method represents a promising and flexible solution for seismic inversion for geologically realistic predictions of multiple properties. The method can be scaled to different grid sizes and adapted to complex field data, balancing between generative deep learning and geostatistical modeling.

## 7. DATA AND MATERIALS AVAILABILITY

The code for SPADE-GANInv and the material used for the inversion case studies are available at https://github.com/romiele/SPADEGAN-for-inversion.

## References

Abdellatif, A., A. H. Elsheikh, D. Busby, and P. Berthet, 2025, Generation of non-stationary stochastic fields using Generative Adversarial Networks: Frontiers in Earth Science, **13**, 1545002.

Amaya, M., N. Linde, and E. Laloy, 2021, Adaptive sequential Monte Carlo for posterior inference and model selection among complex geological priors: Geophysical Journal International, **226**, 1220–1238.

Avseth, P., 2010, Quantitative Seismic Interpretation: Applying Rock Physics Tools to Reduce Interpretation Risk: Cambridge University Press.

Azevedo, L., and A. Soares, 2017, Geostatistical Methods for Reservoir Geophysics: Springer International Publishing.

Azevedo, L., D. Grana, and C. Amaro, 2019, Geostatistical rock physics AVA inversion: Geophysical Journal International, **216**, 1728–1739.

Azevedo, L., D. Grana, and L. De Figueiredo, 2020a, Stochastic perturbation optimization for discrete-continuous inverse problems: Geophysics, **85**, M73–M83.

Azevedo, L., G. Paneiro, A. Santos, and A. Soares, 2020b, Generative adversarial network as a stochastic subsurface model reconstruction: Computational Geosciences, **24**, 1673–1692.

Azevedo, L., R. Nunes, A. Soares, E. C. Mundin, and G. S. Neto, 2015, Integration of well data into geostatistical seismic amplitude variation with angle inversion for facies estimation: Geophysics, **80**, M113–M128.

Buland, A., and H. Omre, 2003, Bayesian linearized AVO inversion: Geophysics, **68**, 185–198.

Caers, J., and T. Hoffman, 2006, The Probability Perturbation Method: A New Look at Bayesian Inverse Modeling: Mathematical Geology, **38**, 81–100.

Chan, S., and A. H. Elsheikh, 2019, Parametric generation of conditional geological realizations using generative neural networks: Computational Geosciences, **23**, 925–952.

Coléou, T., F. Allo, R. Bornard, J. Hamman, and D. Caldwell, 2005, Petrophysical Seismic Inversion: SEG Technical Program Expanded Abstracts 2005, 1355–1358.

Connolly, P. A., and M. J. Hughes, 2016, Stochastic inversion by matching to large numbers of pseudo-wells: Geophysics, **81**, M7–M22.

Cox, M. A. A., and T. F. Cox, 2008, Multidimensional Scaling, *in* Handbook of Data Visualization, Springer Berlin Heidelberg, 315–347.

Curtis, A., and R. Snieder, 1997, Reconditioning inverse problems using the genetic algorithm and revised parameterization: Geophysics, **62**, 1524–1532.

Deutsch, C. V., and A. G. Journel, 1992, GSLIB: Geostatistical Software Library and User's Guide: Oxford University Press.

Dupont, E., T. Zhang, P. Tilke, L. Liang, and W. Bailey, 2018, Generating Realistic Geology Conditioned on Physical Measurements with Generative Adversarial Networks: .

Feng, R., D. Grana, T. Mukerji, and K. Mosegaard, 2022, Application of Bayesian Generative Adversarial Networks to Geological Facies Modeling: Mathematical Geosciences, **54**, 831–855.

Gómez-Hernández, J. J., and X.-H. Wen, 1998, To be or not to be multi-Gaussian? A reflection on stochastic hydrogeology: Advances in Water Resources, **21**, 47–61.

González, E. F., T. Mukerji, and G. Mavko, 2008, Seismic inversion combining rock physics and multiple-point geostatistics: Geophysics, **73**, R11–R21.

Goodfellow, I. J., J. Pouget-Abadie, M. Mirza, B. Xu, D. Warde-Farley, S. Ozair, A. Courville, and Y. Bengio, 2014, Generative Adversarial Networks: .

Grana, D., and E. Della Rossa, 2010, Probabilistic petrophysical-properties estimation integrating statistical rock physics with seismic inversion: Geophysics, **75**, O21–O37.

Grana, D., T. Fjeldstad, and H. Omre, 2017, Bayesian Gaussian Mixture Linear Inversion for Geophysical Inverse Problems: Mathematical Geosciences, **49**, 493–515.

Grana, D., L. Passos De Figueiredo, and L. Azevedo, 2019, Uncertainty quantification in Bayesian inverse problems with model and data dimension reduction: Geophysics, **84**, M15–M24.

Grana, D., L. De Figueiredo, and K. Mosegaard, 2022a, Markov chain Monte Carlo for petrophysical inversion: Geophysics, **87**, M13–M24.

Grana, D., L. Azevedo, L. De Figueiredo, P. Connolly, and T. Mukerji, 2022b, Probabilistic inversion of seismic data for reservoir petrophysical characterization: Review and examples: Geophysics, **87**, M199–M216.

Guardiano, F. B., and R. M. Srivastava, 1993, Multivariate Geostatistics: Beyond Bivariate Moments, *in* A. Soares, ed., Geostatistics Tróia '92, , . Quantitative Geology and GeostatisticsVol. 5. Springer Netherlands, 133–144.

Guo, Q., J. Ba, and C. Luo, 2021, Prestack Seismic Inversion With Data-Driven MRF-Based Regularization: IEEE Transactions on Geoscience and Remote Sensing, **59**, 7122–7136.

Hansen, T. M., K. S. Cordua, and K. Mosegaard, 2012, Inverse problems with non-trivial priors: efficient solution through sequential Gibbs sampling: Computational Geosciences, **16**, 593–611.

He, K., X. Zhang, S. Ren, and J. Sun, 2016, Deep Residual Learning for Image Recognition: 2016 IEEE Conference on Computer Vision and Pattern Recognition (CVPR), 770–778.

Horta, A., and A. Soares, 2010, Direct Sequential Co-simulation with Joint Probability Distributions: Mathematical Geosciences, **42**, 269–292.

Høyer, A.-S., G. Vignoli, T. M. Hansen, L. T. Vu, D. A. Keefer, and F. Jørgensen, 2017, Multiple-point statistical simulation for hydrogeological models: 3-D training image development and conditioning strategies: Hydrology and Earth System Sciences, **21**, 6069–6089.

Ioffe, S., and C. Szegedy, 2015, Batch Normalization: Accelerating Deep Network Training by Reducing Internal Covariate Shift: Proceedings of the 32nd International Conference on Machine Learning, 448–456.

Jia, L., S. Mallick, and C. Wang, 2021, Data-driven prestack-waveform inversion using genetic algorithm: Methodology and examples: Interpretation, **9**, T1065–T1084.

Kingma, D. P., and J. Ba, 2017, Adam: A Method for Stochastic Optimization: ICLR 2015.

Laloy, E., R. Hérault, D. Jacques, and N. Linde, 2018, Training-Image Based Geostatistical Inversion Using a Spatial Generative Adversarial Neural Network: Water Resources Research, **54**, 381–406.

Laloy, E., N. Linde, C. Ruffino, R. Hérault, G. Gasso, and D. Jacques, 2019, Gradient-based deterministic inversion of geophysical data with generative adversarial networks: Is it feasible? Computers & Geosciences, **133**, 104333.

Le Coz, M., J. Bodin, and P. Renard, 2017, On the use of multiple-point statistics to improve groundwater flow modeling in karst aquifers: A case study from the Hydrogeological Experimental Site of Poitiers, France: Journal of Hydrology, **545**, 109–119.

LeCun, Y., Y. Bengio, and G. Hinton, 2015, Deep learning: Nature, **521**, 436–444.

Levy, S., E. Laloy, and N. Linde, 2023, Variational Bayesian inference with complex geostatistical priors using inverse autoregressive flows: Computers & Geosciences, **171**, 105263.

Lim, J. H., and J. C. Ye, 2017, Geometric GAN: .

Linde, N., D. Ginsbourger, J. Irving, F. Nobile, and A. Doucet, 2017, On uncertainty quantification in hydrogeology and hydrogeophysics: Advances in Water Resources, **110**, 166–181.

Lopez-Alvis, J., E. Laloy, F. Nguyen, and T. Hermans, 2021, Deep generative models in inversion: a review and development of a new approach based on a variational autoencoder: Computers & Geosciences, **152**, 104762.

Mariethoz, G., and J. Caers, 2014, Multiple-Point Geostatistics: Stochastic Modeling with Training Images, 1st ed.: Wiley.

Mavko, G., T. Mukerji, and J. Dvorkin, 2020, The Rock Physics Handbook, 3rd ed.: Cambridge University Press.

Melnikova, Y., A. Zunino, K. Lange, K. S. Cordua, and K. Mosegaard, 2015, History Matching Through a Smooth Formulation of Multiple-Point Statistics: Mathematical Geosciences, **47**, 397–416.

Miele, R., and L. Azevedo, 2024, Physics-informed W-Net GAN for the direct stochastic inversion of full-stack seismic data into facies models: Scientific Reports, **14**, 5122.

Miele, R., D. Grana, L. E. Seabra Varella, B. Viola Barreto, and L. Azevedo, 2023, Iterative geostatistical seismic inversion with rock-physics constraints for permeability prediction: Geophysics, **88**, M105–M117.

Miele, R., S. Levy, N. Linde, A. Soares, and L. Azevedo, 2024, Deep generative networks for multivariate full-stack seismic data inversion using inverse autoregressive flows: Computers & Geosciences, **188**, 105622.

Miele, R., B. V. Barreto, P. Yamada, L. E. S. Varella, A. L. Pimentel, J. F. Costa, and L. Azevedo, 2022a, Geostatistical Seismic Rock Physics AVA Inversion With Data-Driven Elastic Properties Update: IEEE Transactions on Geoscience and Remote Sensing, **60**, 1–15.

Miele, R., D. Grana, J. F. Costa, P. Y. Bürkle, L. E. Varella, B. V. Barreto, and L. Azevedo, 2022b, Permeability prediction with geostatistical seismic inversion constrained by rock physics: Second EAGE Conference on Seismic Inversion, 1–5.

Mosegaard, K., and A. Tarantola, 1995, Monte Carlo sampling of solutions to inverse problems: Journal of Geophysical Research: Solid Earth, **100**, 12431–12447.

Mosser, L., O. Dubrule, and M. J. Blunt, 2018, Conditioning of Generative Adversarial Networks for Pore and Reservoir Scale Models: Proceedings.

Nawaz, M. A., and A. Curtis, 2019, Rapid Discriminative Variational Bayesian Inversion of Geophysical Data for the Spatial Distribution of Geological Properties: Journal of Geophysical Research: Solid Earth, **124**, 5867–5887.

Nawaz, M. A., A. Curtis, M. S. Shahraeeni, and C. Gerea, 2020, Variational Bayesian inversion of seismic attributes jointly for geologic facies and petrophysical rock properties: Geophysics, **85**, MR213–MR233.

Nunes, R., A. Soares, L. Azevedo, and P. Pereira, 2017, Geostatistical Seismic Inversion with Direct Sequential Simulation and Co-simulation with Multi-local Distribution Functions: Mathematical Geosciences, **49**, 583–601.

Park, T., M.-Y. Liu, T.-C. Wang, and J.-Y. Zhu, 2019, Semantic Image Synthesis With Spatially-Adaptive Normalization: 2019 IEEE/CVF Conference on Computer Vision and Pattern Recognition (CVPR), 2332–2341.

Rubin, Y., and S. S. Hubbard, 2005, Hydrogeophysics: Springer.

Rwechungura, R. ., E. . Suwartadi, M. . Dadashpour, J. . Kleppe, and B. . Foss, 2010, The Norne Field Case—A Unique Comparative Case Study: All Days, SPE-127538-MS.

Sambridge, M., 1999, Geophysical inversion with a neighbourhood algorithm—I. Searching a parameter space: Geophysical Journal International, **138**, 479–494.

Sambridge, M., and G. Drijkoningen, 1992, Genetic algorithms in seismic waveform inversion: Geophysical Journal International, **109**, 323–342.

Santiago, D. C. C., 2015, 3D Geological Model of the Garn and Not Formations in Norne Field, Mid-offshore Norway: .

Sen, M. K., and P. L. Stoffa, 1991, Simulated annealing, genetic algorithms and seismic waveform inversion: SEG Technical Program Expanded Abstracts 1991, 945–947.

Sen, M. K., and P. L. Stoffa, 2013, Global Optimization Methods in Geophysical Inversion, Second edition.: Cambridge University Press.

Soares, A., 2001, Direct Sequential Simulation and Cosimulation: Mathematical Geology, **33**, 911–926.

Strebelle, S., 2021, Multiple-Point Statistics Simulation Models: Pretty Pictures or Decision-Making Tools? Mathematical Geosciences, **53**, 267–278.

Tang, X., S. Liu, X. Nian, S. Deng, Y. Liu, Q. Ye, Y. Li, Y. Li, T. Yuan, and H. Sun, 2024, Improved adaptive regularization for simulated annealing inversion of transient electromagnetic: Scientific Reports, **14**.

Tarantola, A., 2005, Inverse Problem Theory and Methods for Model Parameter Estimation: Society for Industrial and Applied Mathematics.

Tylor-Jones, T., and L. Azevedo, 2024, A Practical Guide to Seismic Reservoir Characterization: Springer.

Vaswani, A., N. Shazeer, N. Parmar, J. Uszkoreit, L. Jones, A. N. Gomez, Ł. Kaiser, and I. Polosukhin, 2017, Attention is All you Need: 31st Conference on Neural Information Processing Systems (NIPS 2017).

Wang, Z., S. Wang, C. Zhou, and W. Cheng, 2022, Dual Wasserstein generative adversarial network condition: A generative adversarial network-based acoustic impedance inversion method: Geophysics, **87**, R401–R411.

2014, Reservoir Modeling Combining Geostatistics with Markov Chain Monte Carlo Inversion, *in* Lecture Notes in Earth System Sciences, Springer Berlin Heidelberg, 683–687.

2016, Probabilistic Integration of Geo-Information, *in* Geophysical Monograph Series, . Geophysical Monograph Series, 1st ed.Wiley, 93–116.